\documentclass[twocolumn,prl]{revtex4}

\usepackage{graphicx}
\usepackage{dcolumn}
\usepackage{bm}
\usepackage{multirow}
\usepackage{color}
\usepackage[normalem]{ulem}
\usepackage{amsmath}
\usepackage{gensymb}
\usepackage[colorlinks=true]{hyperref}

\definecolor{mblue}{rgb}{0,0.35,0.75}
\definecolor{mgreen}{rgb}{0,0.5,0.5}

\begin{document}
\title{Interlayer excitons in bilayer MoS$_2$ with strong oscillator strength\\ up to room temperature}

\author{Iann C. Gerber$^1$}
\email{igerber@insa-toulouse.fr}
\author{Emmanuel Courtade$^1$}
\author{Shivangi Shree$^1$}
\author{Cedric Robert$^1$}
\author{Takashi Taniguchi$^2$}
\author{Kenji Watanabe$^2$}
\author{Andrea Balocchi$^1$}
\author{Pierre Renucci$^1$}
\author{Delphine Lagarde$^1$}
\author{Xavier Marie$^1$}
\author{Bernhard Urbaszek$^1$}
\email{urbaszek@insa-toulouse.fr}

\affiliation{%
$^1$Universit\'e de Toulouse, INSA-CNRS-UPS, LPCNO, 135 Av. Rangueil, 31077 Toulouse, France}
\affiliation{$^2$National Institute for Materials Science, Tsukuba, Ibaraki 305-0044, Japan}

\begin{abstract}
Coulomb bound electron-hole pairs, excitons, govern the optical properties of semi-conducting transition metal dichalcogenides like MoS$_2$ and WSe$_2$. We study optical transitions at the K-point for \textit{2H} homobilayer MoS$_2$ in Density Functional Theory (DFT) including excitonic effects and compare with reflectivity measurements in high quality samples encapsulated in hexagonal BN. In both calculated and measured spectra we find a strong interlayer exciton transition in energy between A and B intralayer excitons, observable for T$=4 -300$~K, whereas no such transition is observed for the monolayer in the same structure in this energy range. The interlayer excitons consist of an electron localized in one layer and a hole state delocalized over the bilayer, which results in the unusual combination of high oscillator strength and a static dipole moment. We also find signatures of interlayer excitons involving the second highest valence band (B) and compare absorption calculations for different bilayer stackings.  For homotrilayer MoS$_2$ we also observe interlayer excitons and an energy splitting between different intralayer A-excitons originating from the middle and outer layers, respectively.
\end{abstract}

\maketitle
\section{Introduction}
Van der Waals materials have in-plane covalent bonding and the individual layers are held together by the so-called dispersion forces~\cite{Novoselov:2016a,Geim:2013a}. A fascinating aspect of this class of materials is the drastic change of physical properties by changing the sample thickness by just one atomic monolayer. A prominent example is the striking difference between mono- and bilayer graphene \cite{zhang2009direct}. For the van der Waals semiconductor MoS$_2$  the transition from indirect to direct bandgap material occurs when going from bilayers to a monolayer \cite{Mak:2010a, Splendiani:2010a}. These dramatic changes are very different from classical semiconductors like GaAs for example, where the optical properties change gradually with thickness \cite{bastard1982exciton}.\\
\begin{figure}[t]
\includegraphics[width=0.49\textwidth,keepaspectratio=true]{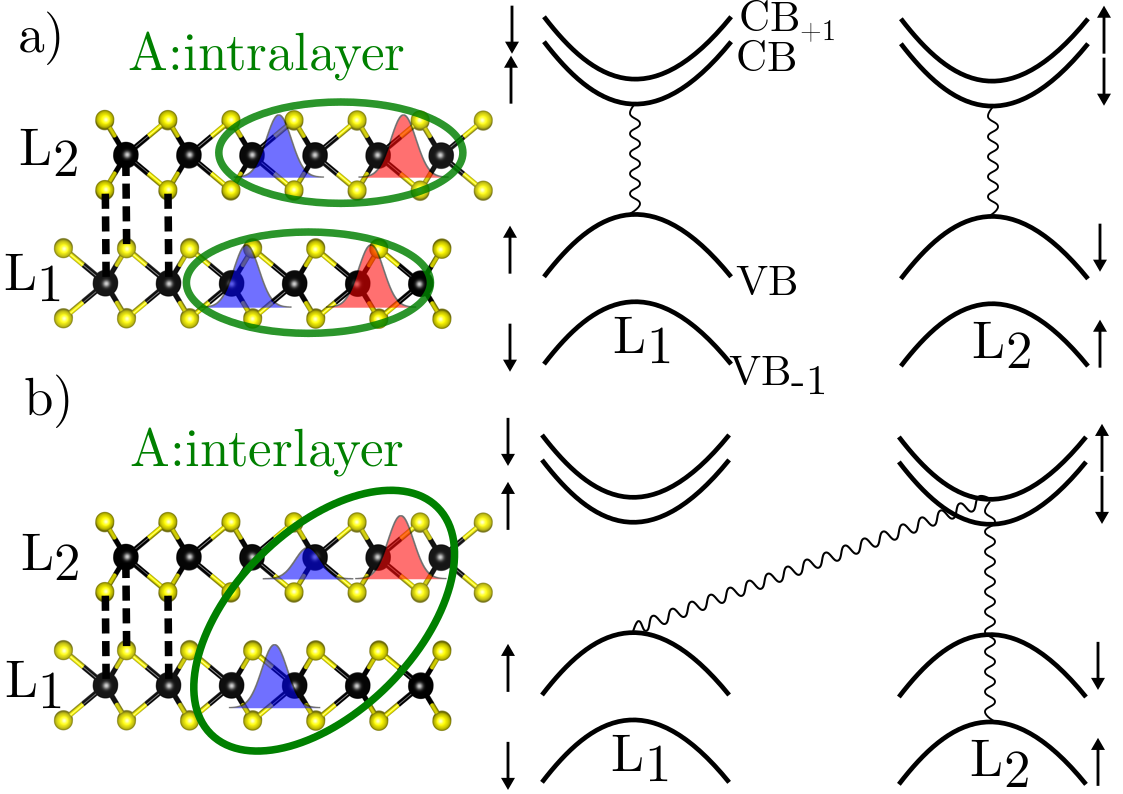}
\caption{\textbf{Schematics of intralayer and interlayer excitons in MoS$_2$ homobilayers.} (a) Intralayer excitons consist of an electron (red) and a hole (blue) in the same layer while in b) an electron localized on one layer interacts with a hybridized hole state to form an interlayer exciton. Optical selection rules,~\textcolor{black}{represented by wavy lines,} for intralayer and interlayer transitions in $k$-space are also given for K points, \textcolor{black}{respecting spin conservation}. For clarity only one interlayer exciton (spin up) is shown, there is also the spin down state with the same energy.}
\label{fig:fig1} 
\end{figure}
\indent The light-matter interaction in monolayer (ML) transition-metal dichalcogenides (TMDs) is governed by Coulomb bound electron-hole pairs, excitons \cite{Wang:2018a,He:2014a,Chernikov:2014a}. As a second layer is added, the light matter interaction is strongly modified since new exciton complexes can form, with electron and hole residing in different layers \cite{gong2013magnetoelectric,Kang:2013b,deilmann2018interlayer,hong2014ultrafast,molas2017optical}, as sketched in Fig.~\ref{fig:fig1}. These interlayer excitons show interesting properties \cite{PhysRevB.97.241404}, also for thicker layers \cite{arora2017interlayer,0022-3719-9-12-029} and more sophisticated van der Waals structures \cite{calman2018indirect}. A very active branch of research investigates spatially indirect interlayer excitons in TMD heterobilayers of with great prospects for spin-valley physics and nano-scale Moire potentials \cite{rivera2016valley,nagler2017giant,2018arXiv180703771T,2018arXiv180904562S,Zhange1601459,yu2017moire,doi:10.1021/acs.nanolett.8b03266}. \\
\begin{figure*}
\includegraphics[width=0.7\textwidth,keepaspectratio=true]{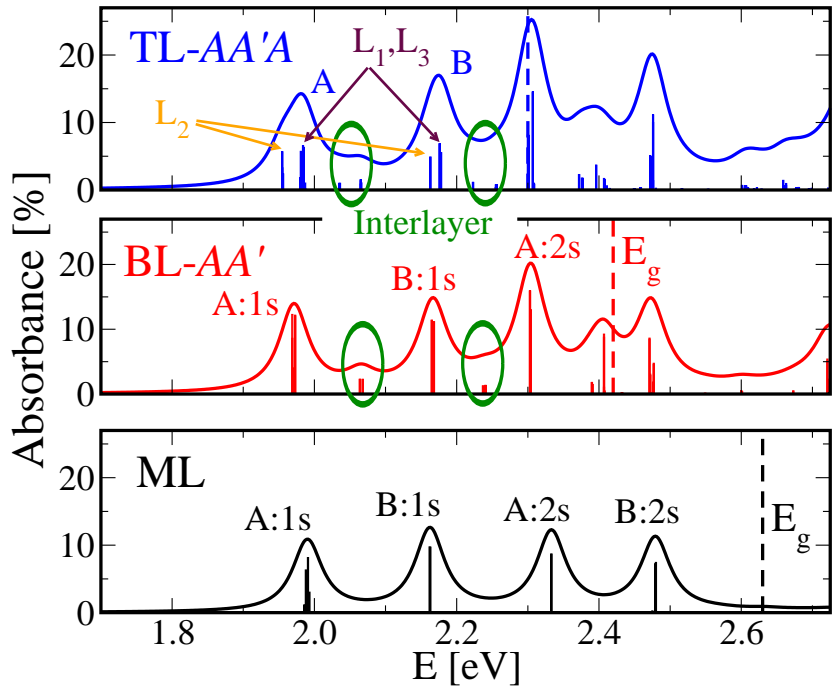}
\caption{\textbf{Calculated absorption spectra for a single mono-, bi and tri-layer.} Interlayer exciton transitions are marked by green circles. Orange arrows for the TL case indicate intralayer transitions involving only the middle layer (L$_2$) of the three layers, while the marroon ones stress the L$_1$, L$_3$ intralayer transitions.See text and appendix for computational details.}
\label{fig:fig2} 
\end{figure*}
\indent In this work we investigate interlayer excitons in homobilayers of MoS$_2$. Contrary to interlayer excitons in TMD heterobilayers, \textcolor{black}{which are indirect both in real and reciprocal space}, we find strong signatures in absorption of the interlayer exciton, about 20 \% of the oscillator strength of the intralayer exciton. Our DFT-$GW$ calculations solving the Bethe-Salpeter equation uncover a strong, spin allowed interlayer exciton peak about 80~meV above the A:1s transition. We find a 20 \% reduction of exciton binding energy of the interlayer exciton compared to the intralayer exciton. Our calculated absorption also predicts an interlayer transition involving the B-valence band located in energy above the B:1s intralayer transition. 
We compare several bilayer stackings in our calculations of optical absorption spectra \cite{gong2013magnetoelectric,PhysRevApplied.4.014002}. Our experiments on high quality bilayer and trilayer MoS$_2$ in hBN show prominent signatures of interlayer excitons up to room temperature in absorption, \textcolor{black}{signalling strong oscillator strength}. The clear manifestation of interlayer excitons opens the way for electric field control of the optical transitions based on their out-of-plane electric dipole \cite{deilmann2018interlayer}. Their strong oscillator strength makes this in addition an interesting system for efficient, tunable coupling to optical microcavities and plasmons \cite{Fogler:2014a,Schneider2018a,low2017polaritons,Liu:2015b,dufferwiel2015exciton}. \textcolor{black}{Our work shows directly the stronger interlayer coupling for the hole states in MoS$_2$ as compared to much weaker coupling expected for K-excitons in WSe$_2$ homobilayers due to the larger A-B valence band separation \cite{Jessi2018,wang2017electrical}. } \\
\indent The paper is organized as follows : First in section \ref{sec:DFTmain} we calculate the bandstructure and optical absorption spectra. Then in section \ref{sec:spectro} we present the corresponding absorption experiments in high quality monolayer, bilayer and trilayer samples. Finally we discuss the comparison between experiment and theory as well as open questions in section \ref{sec:disc}. Computational and experimental details can be found in the Appendices \ref{sec:DFT} and \ref{app:B} .

\section{Band structure and absorption spectra calculations}
\label{sec:DFTmain}

\begin{figure}
\includegraphics[width=0.45\textwidth,keepaspectratio=true]{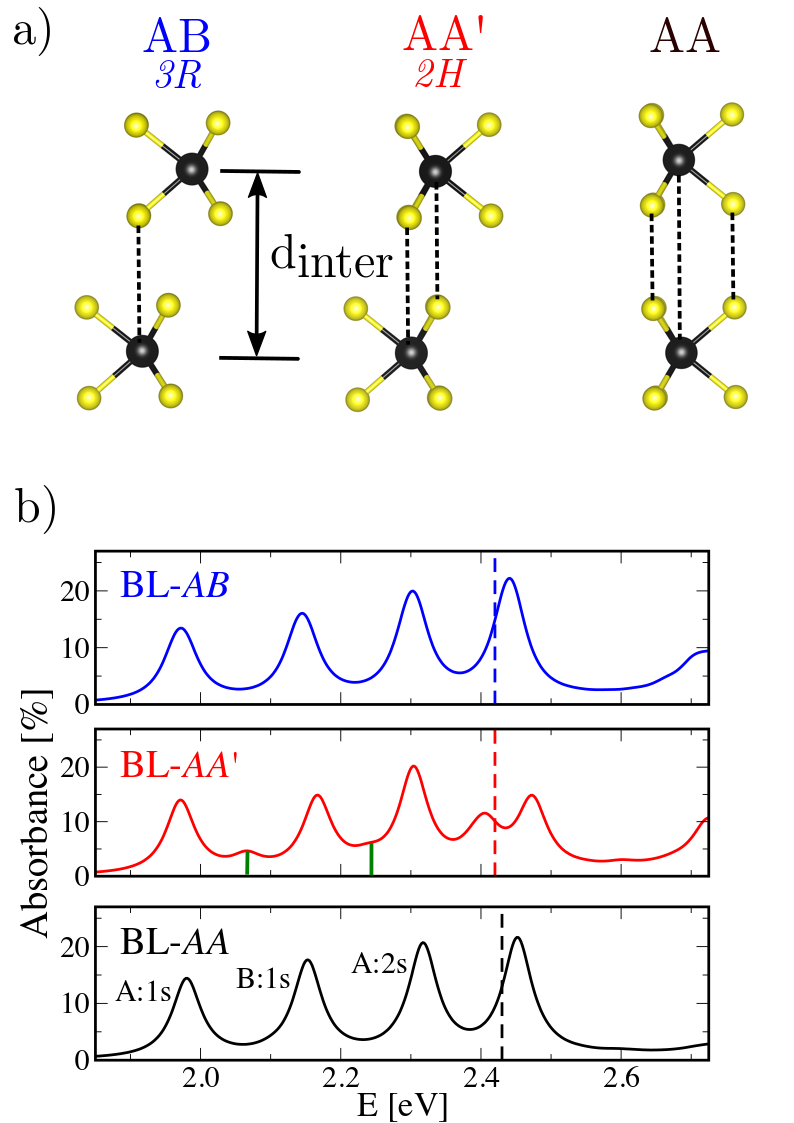}
\caption{\textbf{Calculated absorption spectra for a bilayer, using three different stackings.} The green lines mark interlayer transitions for AA' stackings.}
\label{fig:fig3} 
\end{figure}

The natural MoS$_2$ bilayer (BL) stacking is AA', see Fig.\ref{fig:fig3}a for atomic stacking representation, corresponding to the {\textit{2H}} bulklike symmetry. This is thermodynamically the most stable configuration at the highly accurate Random Phase Approximation level of correlation energy calculations~\cite{He:2014em}. In our case, when using the DFT-D3 exchange-correlation functional scheme of Grimme \textit{et al}~\cite{Grimme:2010ij}, the AB-stacking, prototypical of the {\textit{3R}}-structure \cite{PhysRevApplied.4.014002}, has the same binding energy within meV accuracy, i.e 117 meV/formula unit. In other words, both monolayers gain 117~meV (per elementary cell) by forming a bilayer as compared to staying at infinite distance. AA-stacking is much less favorable: 82 meV per formula unit. The interlayer distance d$_\textrm{inter}$ we find is similar for AA' and AB stacking order, being 6.17 \AA, in good agreement with previous studies~\cite{He:2014em,Liu:1bja}.\\
\indent In Figure \ref{fig:fig2}, we provide absorption spectra calculated from the imaginary part of the complex dielectric function extracted from $GW$+BSE procedure see the appendix for more details, for freestanding ML, BL-AA' and trilayer (TL) in $2H$-like stacking systems, which are the most relevant for samples exfoliated from naturally occurring MoS$_2$. So our calculations for the absorbance are based on a precise determination of the bandstructure and then including the strong excitonic effects. In order to validate our computational approach and precision,~\textcolor{black}{which is of the order of few tens of meV for excitonic peak positions}, we perform calculations for monolayer MoS$_2$ in vacuum and we identify and reproduce the peak positions of the different spectral features as in the work of Qiu \textit{et al.}~\cite{Qiu:2013a,Qiu:2015a}. Note that the 2s feature oscillator strengths are overestimated due to limited number of k-points used in the response function calculations in the BSE step of the calculations. The monolayer results give us confidence for the bilayer system where comparison with experimental data was so far not possible in detail because of the poor optical quality of the structures. As expected, when the number of layers is increased the fundamental gap $E_g$ at the K-point is decreased: 2.62, 2.43 and 2.30~eV for ML, BL and TL respectively, when the multi-layered systems become globally indirect in the $\Gamma$-K direction, as it can be seen in Fig.~\ref{fig:figBS}.\\   
\indent In our calculated AA'-BL absorption spectrum we find an additional transition between the A and B intralayer exciton 1s states, 0.09~eV above the A and B peaks, see Fig.~\ref{fig:fig2}. This peak consists of four degenerate transitions due to spin-splitting and K-K' equivalence. Considering only the spin up in K valley transition as proposed in Fig.\ref{fig:fig1}b), it has 19~\% of the oscillator strength of the corresponding A:1s intralayer transition. Its main contributions come from states corresponding to the valence band spin up: VB of L$_1$ and VB$_{-1}$ of L$_2$ partially hybridized, and a well-localized electron lying in the second lowest conduction band (CB$_{+1}$) states of the other layer. Surprising is the high oscillator strength for this spatially indirect transition, not predicted in earlier work on similar systems \cite{deilmann2018interlayer}. Here the hole states delocalized over the bilayer are important, as the transition we call for brevity interlayer has an intralayer contribution, sketched in Fig.\ref{fig:fig1}b : the intralayer (L$_2$) VB$_{-1}$ to CB$_{+1}$ oscillator strength is roughly 18\% of the spatially indirect VB to CB$_{+1}$ one.  We recall that the symmetry of the first VBs in K are mainly of $d_{x^2-y^2}$ and d$_{xy}$ characters, mixed with $p_{x,y}$ orbitals of S, when the first CBs are made of $d_{z^2}$ orbitals. Interlayer hopping (hybridization) are thus possible in the VBs, helped by the S-$p_z$ orbital contributions in $\Gamma$, but remains impossible for electrons in the CBs \cite{gong2013magnetoelectric,Kormanyos:2015a, PhysRevApplied.4.014002}.\\
\indent Quantitative analysis of the optical transitions in the related system of MoSe$_2$ bilayers in hBN using the Dirac-Bloch equations also predicts an oscillator strength of 20~\% of the interlayer A-exciton compared to the intralayer exciton \cite{PhysRevB.97.241404}. In their work the encapsulation in hBN is taken explicitly taken into account, whereas our calculations are performed in vacuum to avoid high computational cost. Our general target was to see what type of new exciton absorption feature emerges as we go from monolayer to bilayer material - the exact energy position of the transition will be sensitive to screening by the dielectric environment \cite{Stier:2016a,2018arXiv181002130G}. We extract in BL MoS$_2$ the exciton binding energies for the intralayer excitons of about 0.45 eV compared to 0.36 eV for interlayer excitons. This relative comparison shows strong binding for interlayer excitons with carriers residing in different layers, although the absolute values will be smaller in encapsulated samples in hBN principally due to expected band gap renormalization~\cite{2018arXiv181002130G}.\\
\indent For the TL case shown in Fig.~\ref{fig:fig2}, several interesting features are observed: the A:1s state is split, with the intralayer exciton of the central layer (L$_2$) having the largest binding energy and followed in energy by intralayer excitons from the two outside layers (L$_1$,(L$_3$). In our calculations we also see clear signatures of interlayer excitons in TLs.  A set of interlayer transitions is present 0.05~eV above the A peak and again split by 0.03 eV due to the possibility for the carriers to reside either in the central or outside layers. The interlayer exciton oscillator strengths are relatively large as in the bilayer case, around 20\% of the intralayer transitions.\\
\indent Interlayer coupling of VBs and CBs is governed by symmetry and also the spin-orbit splitting between spin-up and spin-down bands, as revealed in very early work in bulk samples \cite{0022-3719-9-12-029}. Whereas interlayer coupling  for electrons is suppressed by symmetry also for AA' stacking \cite{gong2013magnetoelectric}, the interlayer coupling for hole states depends on both symmetry (and more specifically on atomic arrangement between layers) and, if allowed, also on the amplitude of the spin-orbit splitting \cite{gong2013magnetoelectric,PhysRevApplied.4.014002}. In that respect AA' stacking in bilayer MoS$_2$ provides favorable conditions for the observation of interlayer excitons, as the interlayer coupling of VBs is allowed and the spin-orbit splitting is smaller than in MoSe$_2$, MoTe$_2$, WSe$_2$ and WS$_2$. So for sake of completeness, we also calculated the absorption spectra for AA and AB stacking (corresponding to \textit{3R} symmetry in bulk), for which we observe no signature interlayer exciton transitions as shown in the comparison in Fig.\ref{fig:fig3}. 
\begin{figure*}
\includegraphics[width=0.85\textwidth,keepaspectratio=true]{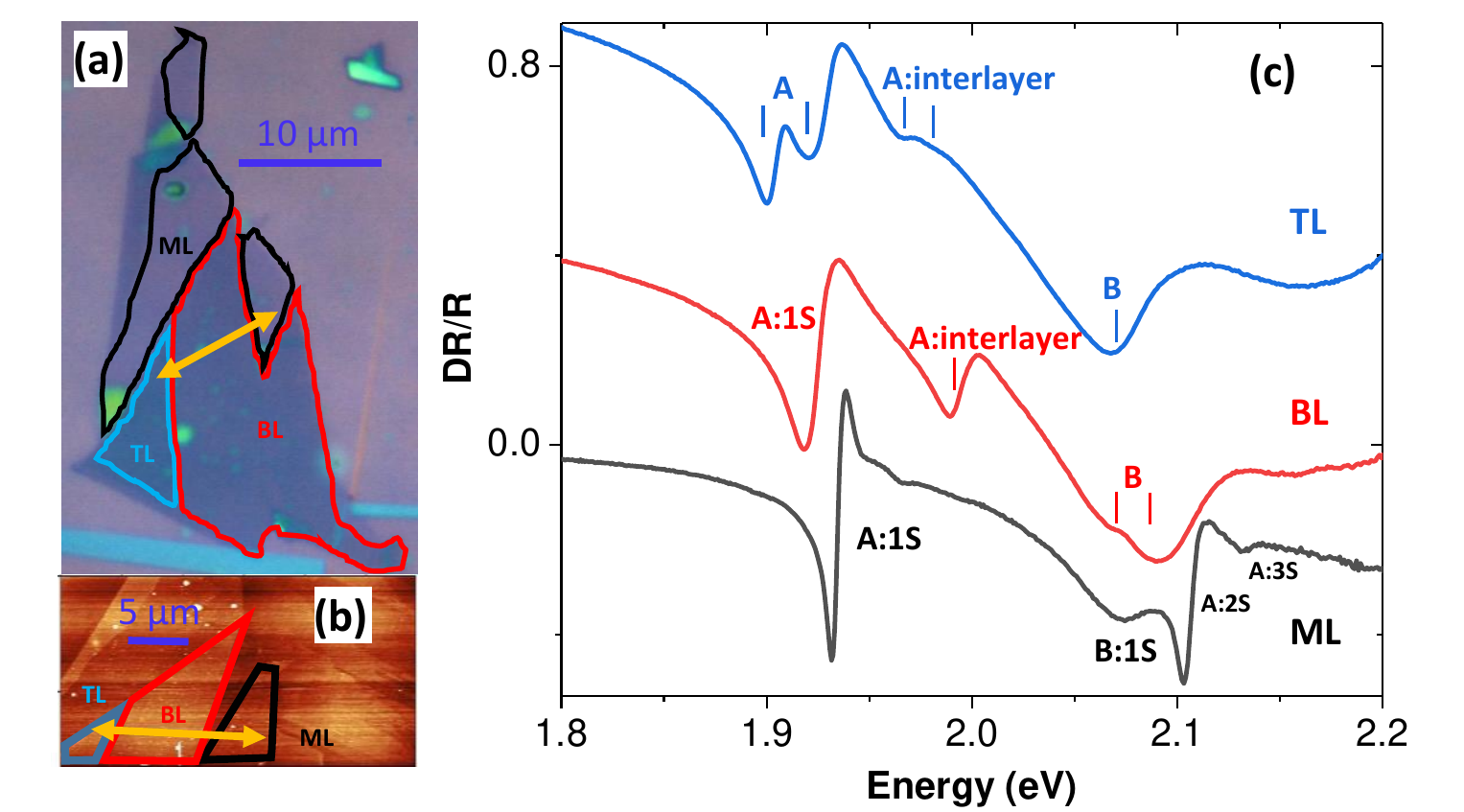}
\caption{\textbf{Monolayer (ML), bilayer (BL) and trilayer (TL) MoS$_2$ encapsulated in hBN } (a) Optical microscope image of the hBN / MoS$_2$ / hBN heterostructure. (b) AFM measurements confirms atomic steps. (c) Differential reflectivity of the three different thicknesses (ML, BL and TL) at a sample temperature of T=4~K, spectra offset for clarity.}
\label{fig:fig4} 
\end{figure*}
\section{Optical Spectroscopy on MoS$_2$ mono-,bi- and trilayers in hBN}
\label{sec:spectro}
Bilayer MoS$_2$ is a fascinating system with tunable properties, explored in a large spectrum of theoretical work \cite{Cheiwchanchamnangij:2012a,bhattacharyya2012semiconductor,deilmann2018interlayer,liu2012tuning,gong2013magnetoelectric,PhysRevB.98.035408} and also experiments \cite{wang2012integrated,wu2013electrical,jiang2014valley,Liu:1bja}.
So far experimental studies of optical properties concentrated on the intralayer exciton. As bilayer MoS$_2$ has an indirect gap, the technique of choice is absorption spectroscopy, either in transmission or reflection geometry, emission in photoluminescence on the other hand is strongly quenched compared to the monolayer \cite{Mak:2010a,Splendiani:2010a}. Further progress was hampered until recently by the very broad optical transition linewidth in MoS$_2$ based nanostructures of the order of 50~meV. Encapsulation in hexagonal boron nitride (hBN) of MoS$_2$ monolayers (MLs) has resulted in considerable narrowing of the exciton transition linewidth down to 1~meV \cite{Cadiz:2017a,Ajayi:2017a} and allowed identification of excited exciton states \cite{Robert:2018a}. This gives access to fine features of the exciton spectra and considerably clearer comparison with theory.
We fabricated a sample with monolayer steps (ML, BL, TL) encapsulated in hBN, see Appendix~\ref{app:B} for details, so we compare all 3 different thicknesses in identical conditions, see Fig.~\ref{fig:fig4}. Atomic force microscopy (AFM) measurements  have been performed in tapping mode, before deposition of the top hBN layer. The topography of Fig.~\ref{fig:fig4}b shows height steps corresponding to monolayer, bilayer and trilayer MoS$_2$.
The extremely different white light reflectivity spectra in Fig.~~\ref{fig:fig4}c are so striking, they can be used for thickness identification, as discussed below. As this sample is exfoliated from \textit{2H} bulk, the bilayer stacking is the thermodynamically most stable AA' configuration, analyzed in detail by DFT in the previous section.
\subsection{Low temperature differential reflectivity}
First we discuss the measurements at low temperature T=4~K. We measure differential reflectivity $(R_\text{ML}-R_\text{sub})/R_\text{sub}$, where $R_{\rm ML}$ is the intensity reflection coefficient of the sample with the MoS$_2$ layer and $R_{\rm sub}$ is the reflection coefficient of the hBN/SiO$_2$ stack. \textcolor{black}{Please note that the overall shape of the differential reflectivity depends on cavity effects (thin layer interference) given by top/bottom hBN and SiO$_2$ thickness. This leads to exciton transition lineshape variations in amplitude and sign in the presented spectra, see \cite{Robert:2018a} for a detailed discussion and comparison with transfer matrix simulations.} \\
\indent \textit{Monolayer.}  As for theory, also for experiment the 1ML sample allows us to validate our approach : the spectra are very similar to the exciton states identified for hBN encapsulated MoS$_2$ in previous work \cite{Robert:2018a}, with a clear signature of the A:2s exciton state superimposed on B:1s, \textcolor{black}{where we find a typical A-B exciton separation of 150~meV in energy \cite{Kormanyos:2015a}}. Here cavity effects determined by the top and bottom hBN thickness used for encapsulation need to be take into account to analyze the oscillator strength \cite{Robert:2018a}. The identification of the A:2s and A:3s as excited A-excitons is confirmed by analyzing the diamagnetic shift in magneto-absorption \cite{Crooker:priv} \textcolor{black}{and using photoluminescence excitation experiments \cite{Robert:2018a}. Note that for the monolayer A:1s to A:2s separation we find an energy of about 170~meV. This is less than the 1s to 2s exciton state separation measured for the B-exciton in uncapped monolayer MoS$_2$ on hBN/SiO$_2$ of about 225~meV \cite{Hill:2015a}. This follows the general trend of finding lower exciton binding energies in hBN encapsulated samples as compared to non-encapsulated ones, underlining the importance of the dielectric environment for the strength of the Coulomb interaction \cite{Stier:2016a,raja2017coulomb}.} \\
\begin{figure*}
\includegraphics[width=0.9\textwidth,keepaspectratio=true]{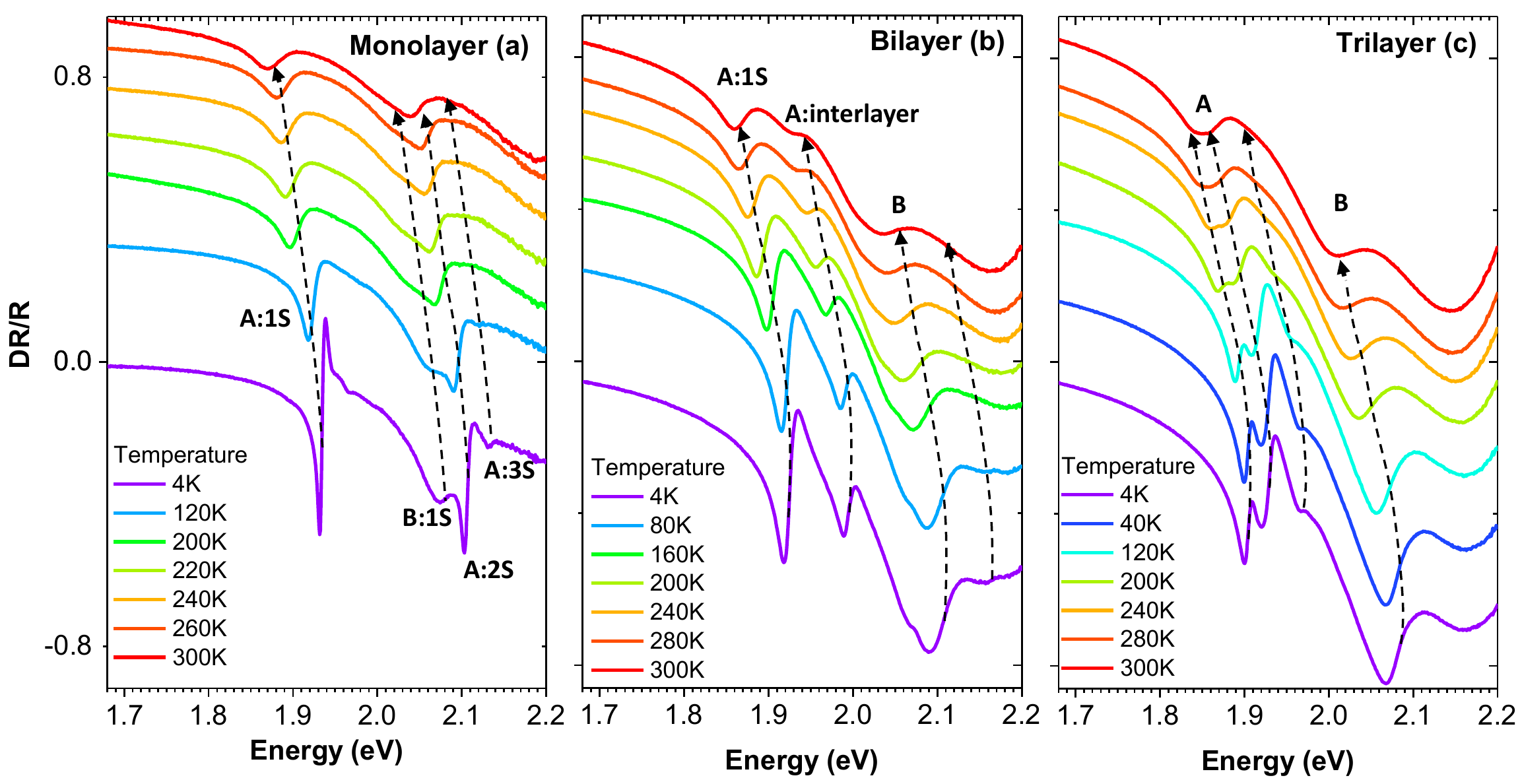}
\caption{\textbf{Optical spectroscopy results:} Temperature dependent differential reflectivity spectra for monolayer (a), bilayer (b) and trilayer (c) MoS$_2$ encapsulated in hBN. }
\label{fig:fig5} 
\end{figure*}
\indent \textit{Bilayer.} The difference between ML and BL absorption is striking : there is an additional transition in Fig.~\ref{fig:fig4}c right between A:1s and B:1s. We attribute this transition 70~meV above the A:1s to the interlayer exciton, with both carriers at the K-point but in different layers. The energy position between A:1s and B:1s fits well with the predictions from our DFT calculations, compare with Fig.~\ref{fig:fig2}. In the region of the B-exciton we find 2 transitions. In addition to B:1s the second peak could be linked for example to the A:2s \cite{Robert:2018a} or the interlayer exciton involving the B-valence band, that we see in the calculated absorption spectrum.\\
\indent \textit{Trilayer.} Finally we investigate a homotrilayer (TL). Here a striking aspect is the observation of not one but two features associated with the main intralayer A-exciton. Here our DFT calculations suggest, see Fig.~\ref{fig:fig2}, that the higher exciton binding energy for intralayer excitons in the middle layer (L$_2$) results in a lower transition energy as compared to the intralayer excitons from the two outer layers (L$_1$,L$_3$). The measured splitting between the two transitions is about 20-25~meV. In addition, between the A- and B-intralayer excitons we observe a feature that we can attribute to interlayer excitons, as we compare our experiment in Fig.~\ref{fig:fig4}c with the calculation in Fig.~\ref{fig:fig2}. Besides a first strong interlayer exciton we also observe a second feature about 15~meV above in energy. From our DFT calculations for the interlayer exciton different energies are expected for carriers residing in the inner or outer layers, similar to the splitting observed in the intralayer exciton case.\\
\indent \textcolor{black}{Both homobilayer and trilayers are indirect semiconductors. The optical transitions involving excitons direct in momentum space with carriers from the K-points can be broadened compared to the ML case due to relaxation towards the lower lying indirect bandgap.}
\subsection{Temperature evolution of absorption from 4~K to room temperature}
\indent In Fig.~\ref{fig:fig5} we analyze the temperature evolution of the differential reflectivity spectra. The evolution of the ML spectrum up to room temperature shows a standard shift of the A:1s transition with temperature.  At 300 K at first glance two strong transitions are visible, separated by 170~meV. The B:1s absorption is very broad and its energy cannot be fitted precisely. Surprisingly, the most pronounced feature at higher energy is not the B:1s but the A:2s state. In general to observe excited exciton states also at room temperature is consistent with the high binding energy of these intralayer excitons in hBN encapsulated MoS$_2$ of about 220~meV \cite{Robert:2018a}. \\
\indent Remarkably, for the bilayer in Fig.~\ref{fig:fig5}b the intralayer but also the interlayer transition are still observable at room temperature, again consistent with a high exciton binding energy, as indicated by our {\textit{ab initio}} results. For the trilayer, the double feature for the intralayer A-exciton is visible at all temperatures. The main interlayer exciton is still discernible at 240~K.
\section{Discussion}
\label{sec:disc}
For simplicity, in our theory and experiment we concentrate on optical transitions with large oscillator strength, so in our optical absorption spectra we have no clear signature of possible optical transitions indirect in k-space, for example, that involve carriers from the $\Gamma$-point \cite{Zhao:2013c,Kormanyos:2013a}. In this work we exclusively discuss transitions involving carriers in different layers and bands, but all at the K-point. The band structure of TMD bilayers is rather complex at the K-point, with spin-split conduction and valence bands \cite{Kosmider:2013b,Liu:2013a,gong2013magnetoelectric}. Already in a single particle picture this gives rise to several optical transitions. In optical spectroscopy experiments we work with excitons not band to band transitions, so all energy scales are renormalized by the Coulomb interaction, the direct and exchange terms. \\
\indent We now try to analyze why the interlayer and intralayer A-excitons have different transition energies, similar arguments hold for the B-excitons. 
In the bilayer absorption measurements in Figs.~\ref{fig:fig4}c and \ref{fig:fig5}b and also calculations in Fig.~\ref{fig:fig2} we observe the intralayer exciton transition about $\Delta_{\text{exp}}=$70~meV lower in energy than the interlayer exciton. Several effects can contribute to this difference :\\
(i) Difference in intralayer (calculated 0.45 eV) and interlayer exciton binding energy (0.36~eV). Although the calculations are for structures in vacuum and our sample is encapsulated in hBN, we see this difference is significant and will provide an important contribution to $\Delta_{\text{exp}}$. The physical origin of the difference in binding energies can come from the different spatial extension of the exciton in the intra- and interlayer configuration. The effective mass for spin-up and spin-down conduction and valence bands that we can extract from our band structure calculations is another source for differences in the binding energies of different exciton species \cite{Kormanyos:2015a}. For the A-interlayer exciton the difference in mass of the two lowest lying conduction bands is relevant, and we find  0.47~$m_0$ for CB$_{+1}$ and 0.42~$m_0$ for CB, respectively, where $m_0$ is the free electron mass. Though significant, our calculations show this mass difference remains a smaller contribution compared to the exciton spatial extension change between the two configurations.\\
(ii) Due to spin-conservation in optical dipole transitions, the interlayer excitons is formed with an electron in the second lowest, not lowest conduction band, see Fig.~\ref{fig:fig1}b. The conduction band spin splitting is estimated to be in the meV range~\cite{Kosmider:2013b,Liu:2013a}, we find 13~meV in our calculations, see Fig.~\ref{fig:figBS}, in very good agreement with a recent experimental measurement~\cite{Pisoni:2018vn}. So this conduction band spin splitting can contribute to $\Delta_{\text{exp}}$, but is not the dominating term. \\
(iii) The exchange terms of the Coulomb interaction are also important and for the case of MoS$_2$ might reverse the order in energy of the spin-allowed and spin forbidden transitions \cite{Echeverry:2016,zhang2017magnetic}. \\
\indent In a very recent preprint interlayer excitons in MoS$_2$ \cite{2018arXiv181000623S} are discussed in detail for bilayer and trilayers using k.p theory and comparing with magneto-optics. Although the theoretical approach is very different from our ab-initio calculations, both approaches agree on the existence and importance of interlayer K-point excitons. Predicting the exact energy positions is still challenging due to the uncertainties in amplitude and sign of the conduction band spin splitting \cite{molas2017brightening}, the Coulomb exchange terms and also the effective masses \cite{Pisoni:2018vn}, see \cite{2018arXiv181000623S} for a complementary analysis.  \\
\indent \textit{In conclusion,} interlayer excitons with high oscillator strength are found in post-DFT calculations and optical absorption measurements on MoS$_2$ homobilayers and trilayers. Their optical signatures are visible up to room temperature. The interlayer excitons involve an electron in one layer and a hole delocalized over both layers. This combines in principle large oscillator strength with a large static dipole moment, which is a desirable configuration for coupling quantum tunneling with cavity photons, previously reported at cryogenic temperatures in III-V semiconductor nano-structures \cite{Cristofolini704}.

\begin{acknowledgements}  
We thank Alexandre Pierrot and Benjamin Lassagne for technical assistance with AFM measurements. We thank Kristian Thygesen, Clement Faugeras, Misha Glazov and Atac Imamoglu for stimulating discussions. 
We acknowledge funding from ANR 2D-vdW-Spin, ANR VallEx, Labex NEXT projects VWspin and MILO, ITN Spin-NANO Marie Sklodowska-Curie grant agreement No 676108 and ITN 4PHOTON Nr. 721394. X.M. also acknowledges the Institut Universitaire de France. Growth of hexagonal boron nitride crystals was supported by the
Elemental Strategy Initiative conducted by the MEXT, Japan and the CREST
(JPMJCR15F3), JST. Finally, I. C. Gerber thanks the CALMIP initiative for the generous allocation of computational times, through the project  p0812, as well as the GENCI-CINES and GENCI-IDRIS for the grant A004096649.
\end{acknowledgements}


\appendix

        \setcounter{figure}{0}
        \renewcommand{\thefigure}{S\arabic{figure}}%
        
\section{Computational Details}
\label{sec:DFT}
\indent The atomic structures, the quasi-particle band structures and optical spectra are obtained from DFT calculations using the VASP 
package \cite{Kresse:1993a,Kresse:1996a}. It uses the plane-augmented wave scheme \cite{blochl:prb:94,kresse:prb:99} to treat core electrons. 
Perdew-Burke-Ernzerhof (PBE) functional \cite{Perdew:1996a} is used as an approximation of the exchange-correlation electronic term, to build the starting wave-function for $GW$ calculations.  
During geometry's optimization step of all the hetero-structures, performed at the PBE-D3 level~\cite{Grimme:2010ij}, all the atoms were allowed to relax with a force convergence criterion below $0.005$ eV/\AA,  in order to include van der Waals interaction between layers. The optimized lattice parameter of MoS$_2$, obtained at the PBE level, used for all the calculations is 3.22 \AA .
 A grid of 15$\times$15$\times$1 k-points has been used, in conjunction with a vacuum height of  21.9~\AA, for all the calculation cells, to take benefit of error's cancellation in the band gap estimates~\cite{Huser:2013fs}, 
and to provide absorption spectra in good agreement with experiments~\cite{Klots:2014a, MolinaSanchez:2013}. An energy cutoff of 400 eV and a gaussian smearing of 0.05 eV width have been chosen for partial occupancies, when a tight electronic minimization tolerance of $10^{-8}$ eV is set to determine with a good precision the corresponding derivative of the orbitals with respect to $k$ needed in quasi-particle band structure calculations. 
Spin-Orbit Coupling (SOC) was also included non-self-consistently to determine eigenvalues and wave functions as input for the full-frequency-dependent $GW$ 
calculations~\cite{Shishkin:2006a} performed at the $G_3W_0$ level ~\cite{Echeverry:2016}. The total number of states included in the $GW$ procedure is set to 1280, in conjunction with an energy cutoff of 100 eV for the response function, after a careful check of the direct band gap convergence (smaller than 0.1 eV as a function of k-points sampling). Band structures have been obtained after a Wannier interpolation procedure performed by the WANNIER90 program~\cite{Mostofi:2008ff}.
All optical excitonic transitions have been calculated by solving the Bethe-Salpeter Equation~\cite{Hanke:1979to,Rohlfing:1998vb}, using the twelve highest valence bands and the sixteen lowest conduction bands to obtain eigenvalues and oscillator strengths on all systems. From these calculations, we report the absorbance values by using the imaginary part of the complex dielectric function.

\begin{figure}
\includegraphics[width=0.48\textwidth,keepaspectratio=true]{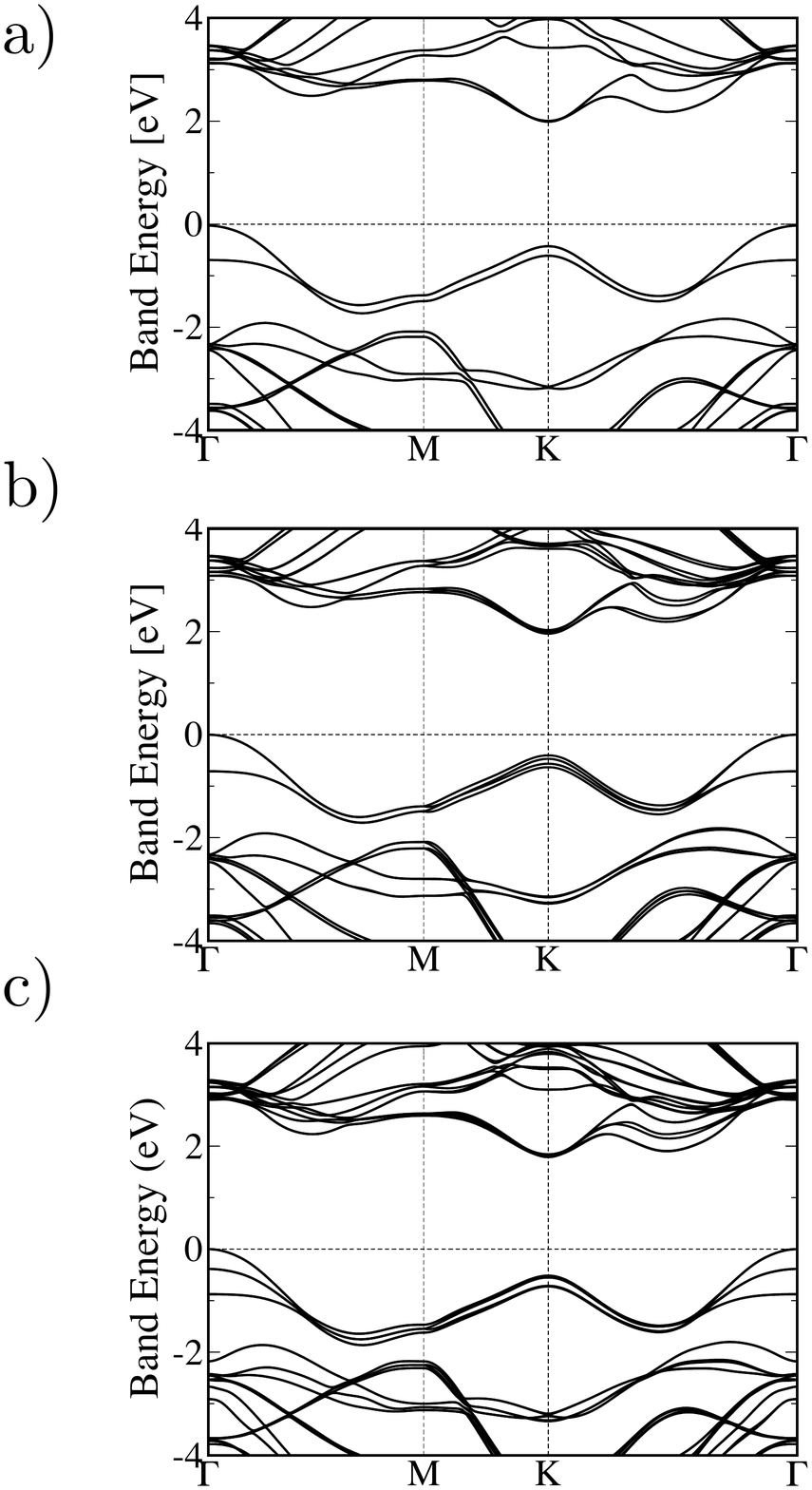}
\caption{{\textbf{Quasiparticle band structures}} at the $G_3W_0$ level in a) AA' BL, b) AB  BL and c) AA'A TL stacking.}
\label{fig:figBS} 
\end{figure}
        
\section{Experimental Methods}\label{app:B}
The samples are fabricated by mechanical exfoliation of bulk MoS$_2$ (commercially available from 2D bulk semiconductors) and very high quality hBN crystals on 83-nm SiO$_2$ \cite{Taniguchi:2007a} on a Si substrate.  The experiments are carried out in a confocal microscope built in a vibration free, closed cycle cryostat with variable temperature. The excitation/detection spot diameter is $\sim  1 \ \mu$m. Reflectivity measurements were performed with a power-stabilized white halogen lamp for sample temperatures T= 4 - 300 K. The  reflectivity  signal  is  dispersed  in  a  spectrometer  and  detected  with  a  Si-CCD camera \cite{Robert:2018a}.

\end{document}